\documentclass[sigchi-a]{acmart}
\usepackage{balance}       % to better equalize the last page
\usepackage{graphics}      % for EPS, load graphicx instead 
\usepackage{url}      % llt: nicely formatted URLs

\usepackage{csquotes}
\usepackage{enumitem}
\usepackage[linguistics]{forest} % decision tree

\usepackage{color}
\usepackage{siunitx}
\newcolumntype{L}[1]{>{\raggedright\let\newline\\\arraybackslash\hspace{0pt}}m{#1}}
\newcolumntype{C}[1]{>{\centering\let\newline\\\arraybackslash\hspace{0pt}}m{#1}}
\newcolumntype{R}[1]{>{\raggedleft\let\newline\\\arraybackslash\hspace{0pt}}m{#1}}

  {\list{}{\leftmargin=0.15in\rightmargin=0.15in}\item[]}%
  {\endlist}
  
\makeatletter
\renewcommand\subsubsection{\@startsection{subsubsection}{3}{\z@}%
  {-.75\baselineskip \@plus -2\p@ \@minus -.2\p@}%
  {.1\linespacing}%
  {\@subsubsecfont}}
   
\def\@subsubsecfont{\sffamily} 
\makeatother

\usepackage[titles]{tocloft}
\def\BibTeX{{\rm B\kern-.05em{\sc i\kern-.025em b}\kern-.08emT\kern-.1667em\lower.7ex\hbox{E}\kern-.125emX}}
    
\copyrightyear{2019}
\acmYear{2019}
\setcopyright{acmlicensed}
\acmConference[CHI 2019 HCML Workshop]{CHI Conference on Human Factors in Computing Systems Proceedings}{May 4--9, 2019}{Glasgow, Scotland Uk}
\acmBooktitle{Human-Centered Machine Learning Perspectives Workshop at CHI Conference on Human Factors in Computing Systems Proceedings (CHI 2019) , May 4--9, 2019, Glasgow, Scotland Uk}

\begin{document}

%
% The "title" command has an optional parameter, allowing the author to define a "short title" to be used in page headers.
\title{Two Case Studies of Experience Prototyping Machine Learning Systems in the Wild}

%
% The "author" command and its associated commands are used to define the authors and their affiliations.
% Of note is the shared affiliation of the first two authors, and the "authornote" and "authornotemark" commands
% used to denote shared contribution to the research.
\author{Qian Yang}
\affiliation{%
  \institution{Carnegie Mellon University}
}
\email{yangqian@cmu.edu}

%
% By default, the full list of authors will be used in the page headers. Often, this list is too long, and will overlap
% other information printed in the page headers. This command allows the author to define a more concise list
% of authors' names for this purpose.
\renewcommand{\shortauthors}{Q. Yang}

\maketitle

\section{Introduction}

Throughout the course of my Ph.D., I have been designing the UX of various machine learning (ML) systems.
%, situating them into various contexts in the real world, and studying how users interact with them. These systems range from mobile crowd-sourcing systems, autonomous cars, clinical ML, and more.
In this workshop, I share two projects as case studies in which people engage with ML in much more complicated, and nuanced ways than the technical HCML work might assume. 

The first case study describes how cardiology teams in three hospitals used a clinical decision-support system that helps them decide whether and when to implant an artificial heart to a heart failure patient \cite{yangVAD2016,yang-vad-unremarkable}. I demonstrate that physicians cannot draw on their decision-making experience by seeing only patient data on paper. They are also confused by some fundamental premises upon which ML operates. For example, physicians asked: Are ML predictions made based on clinicians' best efforts? %Is it ethical to make decisions based on previous patients' collective outcomes?
In the second case study, my collaborators and I designed an intelligent text editor, with the goal of improving authors' writing experience with NLP (Natural Language Processing) technologies \cite{sketching-NLP}. We prototyped a number of generative functionalities where the system provides phrase-or-sentence-level writing suggestions upon user request. When writing with the prototype, however, authors shared that they need to ``see where the sentence is going two paragraphs later" in order to decide whether the suggestion aligns with their writing; Some even considered adopting machine suggestions as plagiarism, therefore ``\textit{is simply wrong}".

By sharing these unexpected and intriguing responses from these real-world ML users, I hope to start a discussion about such previously-unknown complexities and nuances of -- as the workshop proposal states -- ``\textit{putting ML at the service of people in a way that is accessible, useful, and trustworthy to all}".

\section{Two Case Studies}

\subsection{Cardiologists Using Prognostic Decision Support Systems}

Clinical decision support tools (DST) promise improved healthcare outcomes by offering data-driven insights. Interestingly, almost all these tools have failed when migrating from research labs to clinical practice in the past 30 years \cite{Musen2014clinical}.

We are collaborating with biomedical researchers on the design of a DST supporting the decision to implant an artificial heart. The artificial heart, VAD (ventricular assist device), is an implantable electro-mechanical device used to partially replace heart function. For many end-stage heart failure patients who are not eligible for or able to receive a heart transplant, VADs offer the only chance to extend their lives. Unfortunately, many patients who received VADs die shortly after the implant. In this light, a DST that can predict the likely trajectory a patient will take post-implant, should help identify the patients who are mostly likely to benefit from the therapy.

Our previous study investigating the VAD decision processes \cite{yangVAD2016} revealed that clinicians, for most cases, did not find the implant decision challenging; thus, they had no desire for computational support. In addition, the extremely hierarchical workplace culture stratified senior physicians who make implant decisions and the mid-level clinicians who use computers. Almost no VAD decision-making took place in front of a computer.

We designed a radically new DST that automatically generates slides for the required decision meetings which all clinicians attend. The design embeds ML prognostic decision supports into the corner of their meeting slides. We wanted decision makers to encounter the computational advice at a relevant time and place across the decision process, and we wanted this support to only slow them down for the few cases where the DST adds value to the decision. 

\begin{marginfigure}
    \centering
    \includegraphics[width=1.3\linewidth]{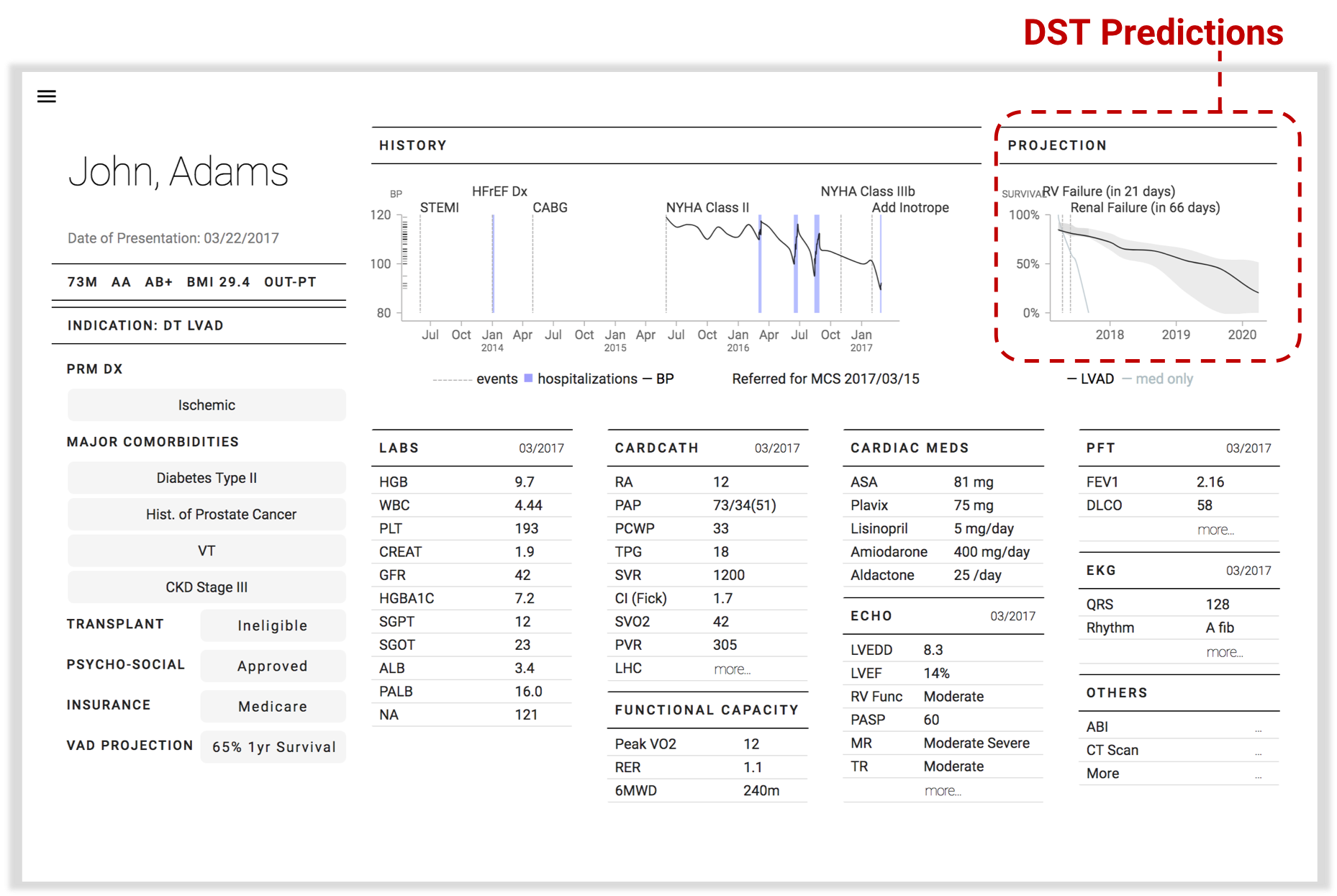}
    \caption{The decision meeting slide design. We designed a DST that automatically generates decision-meeting slides for clinicians with subtly embedded machine prognostics at the top right corner.}
    \vspace{0.05cm}
    \label{fig:screenshot}
\end{marginfigure}

In the field evaluation of this DST design, VAD care teams -- including cardiologists, surgeons, nurse practitioners, social workers, and more -- at three VAD implant hospitals across the U.S. used the tool in their implant meetings.
Clinicians' responses suggest that clinicians are more likely to encounter and embrace a DST that binds ``unremarkable" decision supports with their current work routine. More importantly, once unblocked by attitudinal resistance, clinicians' responses depicted many un-expected complexities and intricacies in terms of what they really desire from ML decision supports. Below are few examples.

\subsubsection{Challenges of Engaging Patient Cases via Data}
Clinicians shared that they could not draw on their experience of making critical clinical decisions seeing only patient data on paper. Physicians described the meeting data as merely a surrogate for the actual patient. The data did not allow them to see patients ``\textit{as a whole.}'' They stressed that to understand a patient clinically, they needed to ``\textit{look at the patient, talk to the patient, take care of the patient.}'' Social workers could not comment on the patient cases because that they ``\textit{had not met with this patient nor talked to their family}". 

\begin{sidebar}
\textit{A very sick but highly motivated patient can do better than their illness would otherwise be left them, compared to a less sick, less motivated patient. These things are hard to capture. The eyeball tests.} (Surgeon, B6)
\end{sidebar}

Interestingly, clinicians also had wildly different readings into the same DST prognostics. We presented the same two synthetic cases with the same implant survival predictions to all participants. They reacted to and interpreted the cases in wildly different ways. Some viewed the survival estimate as implying that an implant would not work. ``\textit{Gee...~VAD is futile here.}'' Others viewed the DST output as implying the patient should be immediately implanted, before things got worse. `\textit{`We still have a chance.}'' Few clinicians believed that all VAD implant candidates would have a similar prognosis as the synthetic case we presented: ``\textit{This chart is meaningless.}'' 

\subsubsection{Are ML Prognostics Facts OR Predictions?}
\begin{sidebar}
\textit{I think if you continue to call it ``VAD projections'' 65\%, people are going to poke holes at it. They are gonna try to prove you wrong. This [DST projection] is just what the historical outcomes were. But this guy is different, this guy has his own things that make him special.} (Collaborating cardiologist, hospital A)
\end{sidebar}

Clinicians frequently asked us to clarify whether DST prognostics are predictions that carry agency and subjectivity, or if predictions are facts rooted in historic data. Some voiced strong concerns that applying ``\textit{populational statistics}'' to individual patient decision making was unethical.

\subsubsection{Are ML Predictions Based on Clinicians' Best Efforts?}

\begin{sidebar}
``\textit{If we think that we will be able to tell everybody what to do based on a model, we ignore the fact that we also have tools and mechanisms for dealing with the uncertainty that is inherent when putting VADs in patients.}'' (Cardiologist)
\end{sidebar}

``\textit{These predictions are (what will happen) despite our best efforts, right?}" -- Many clinicians' questions, as well as their discussion around the DSTs, revealed a tension between what they saw as the DST's static view of patient conditions and the clinicians' desire and ability to also focus on future actions and interventions. They wanted to know which modifiable factors most influenced the DST predictions. They wanted to be able to offer treatments that they could improve these factors, thus increasing the likelihood of a positive surgical outcome at some time in the future.

\subsubsection{What Does ``Now'' Mean in ML Predictions?}
Our DST visualized the patient outcome predictions, including life expectancy, estimated time until right heart failure, and likely cause of death.
Clinicians were very confused by this notion of ``now'' because it was extremely unlikely that they would implant a patient on the same day as the decision meeting.  ``\textit{Is that 21 days (life expectancy) from today? If we are gonna lose the patient in 21 days [21 days following after implant], can we just wait?}''

\subsection{Writers Using Intelligent Text Editors}

In this project, I collaborated with a group of NLP researchers on designing intelligent functionality offerings in a Word document editor.
Prior HCI research has utilized NLP for providing writing assistance in several ways, for example, suggesting next sentences as inspiration \cite{lexical-writing-assistant,writing-novel-with-machine-UW}.

Our goal was to improve individual users' writing \textit{experience}. This focus on \textit{experience} means that our focus was NOT on whether or how authors needed to produce ``better" writing \textit{products}. Instead, we wanted understand how authors themselves want to be supported by machine intelligence in the \textit{process} of writing. Relatedly, we did not assume that authors in need of assistance, because our very design task was to search for the writing contexts in which authors are in desperate need of assistance such that they would embrace the likely imperfect ML-generated writing suggestions.

% First study
First, we conducted a field study of 18 participants to understand their needs and wants in writing. We invited them to record their screen for 40 minutes as they were writing \textit{one of their own documents}. Participants then walked us through their thought process in writing during the time of screen recording and discussed unmet needs for writing assistance. 
Accordingly, we generated a prioritized set of intelligent function offering ideas. These ideas include search-based, targeted rephrasing, suggesting references, citations and examples for a selected content, etc.
% For example, we envisioned an ask-your-reader function that compares a user's writing with their target venues', helping them to account for readers' likely expectations. We created prototypes of these early ideas and tested them in a second user study.
We then turned to building prototypes in order to rapidly experiment on these ideas with users. We wanted to test the ideal behavior of our envisioned intelligent assistance with users to see if we were pursing the right design direction; We also wanted to probe users' reactions to a more realistic range of NLP-powered behaviors and errors to account for these reactions and expectations when improving on our design.

\begin{marginfigure}
\centering
 \includegraphics[width=\linewidth]{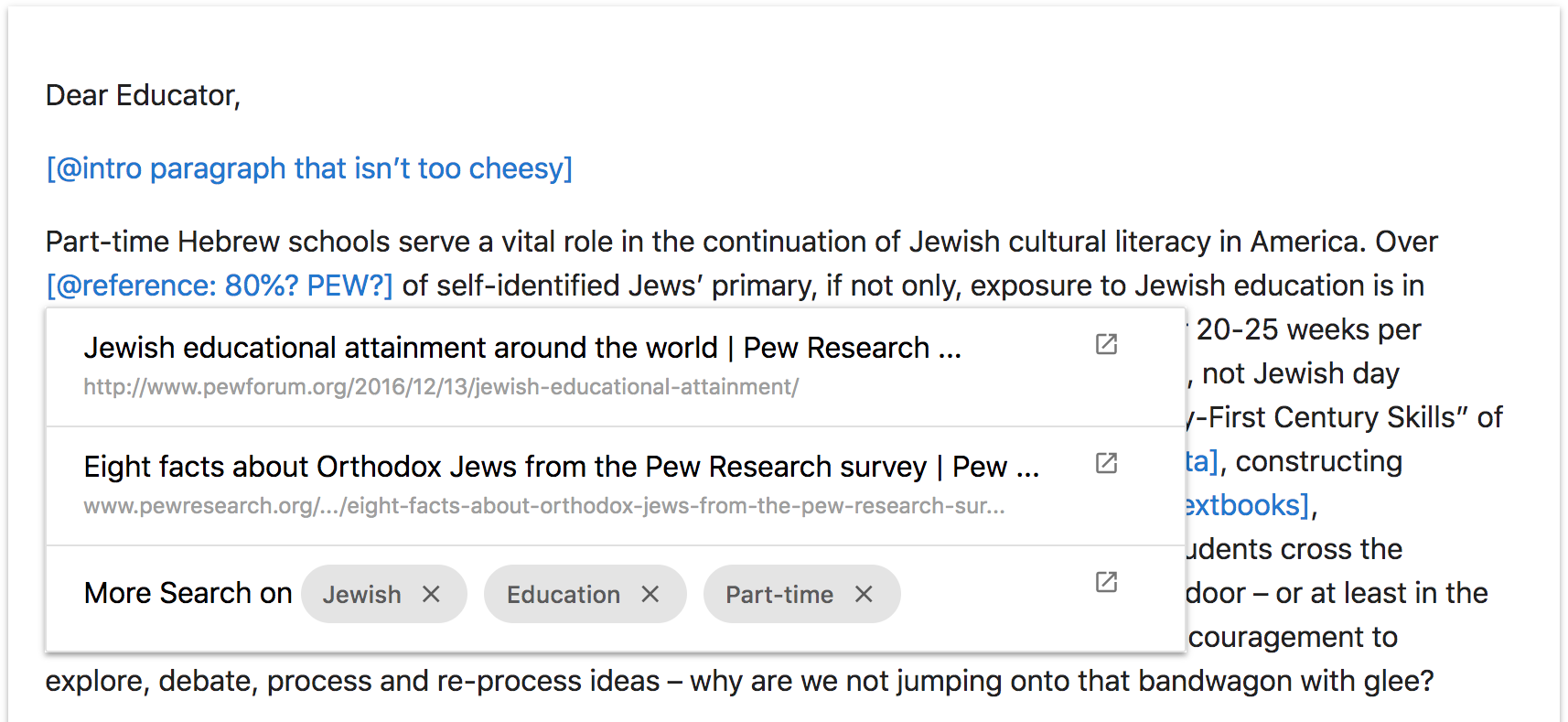}
 \caption{The intelligent text editor prototype. At any time of their writing, users type @ to signal the start of an intelligent function request and Enter to end. %A tooltip suggested our available function offerings when participants typed any request, though participants were free to type anything as request. 
 When they click on a request, intelligent assistance pops out. This prototype probes users' needs and wants for writing assistance, and their reactions to the simulated intelligent responses.
 }
 \label{fig:probe}
\end{marginfigure}

To best simulate the likely errors of generative NLP systems, we designed different hybrids of WoZ and off-the-shelf toolkits for each NLP-powered interaction design. Each hybrid was designed to mimic the likely architecture of its underlying generative neural networks. For example, we prototyped the generative writing assistance functionality with a simulator composed of multiple wizards and a meta classifier. When a user study participant request a piece of machine-generated text, one wizard produces a topically relevant response; the second wizard takes charge of the response fluency; the third focuses on the coherence between the generated text and the writers; the forth adds domain knowledge to the response, the fifth generates random words, and so on. The meta classifier then assembles all wizard's responses into the final response returned to the user. 
% These wizards' roles were designed based on common architectures of generative neural networks. 
We simulate different kinds/degrees of generative errors by tuning the weights that each wizard carry. As such, we could probe user study participants on their preference among various designs of generative writing assistance as well as their error tolerance. 

In the second user study, we invited the 18 participants to use the prototypes as they were writing one of their own documents. This is different from almost all previous HCI work on writing assistance systems that we are aware of, in which researchers typically invited participants to write on pre-determined topics for a particular time duration (e.g., as in \cite{lexical-writing-assistant,writing-novel-with-machine-UW}). 
Our prototypes again triggered unexpected reactions, ones that are distinctively different from either previous HCI studies or what the participants verbally described as desirable from an intelligent text editor. 

\subsubsection{Is Adopting Machine Generated Writing Plagiarism?}
One of the functionalities offered in the prototype is generating sentence-or-phrase-level writing suggestions. For example, when participants type ``@ add an opening with a quote'' (@ signals a request for intelligent assistance) and click on the request, the prototype surfaces a list of opening sentence suggestions. Participants in the initial user study expressed a desire for such functionalities for they can save writers' efforts to search for relevant quotes, examples, or references online and to integrate into their own writings.

\begin{sidebar}
``\textit{Even if I just liked this scaffolding (in the machine-generated suggestion), I wouldn't take these exact words. It's like... In my school, five or six consecutive words from any other piece of media that isn't referenced as a quote are considered plagiarism. People get spelled from school [...] It's a societal judgemental thing.}" (P9)
\end{sidebar}

Interestingly, when participants saw the machine-generated suggestions to their own writing, they instantly became more resistant and expressed a much stronger sense of ownership or their writing than they had initially expressed. ``\textit{Isn't this plagiarism?}" Few participants referred to adopting a machine suggestion directly as ``\textit{stealing a sentence from another article}" and ``\textit{it just feels wrong!}"

Instead of accepting machine-suggested sentences, most participants browsed many, many suggestions, and from different suggestions picked parts of the sentences, semantics, word choices or references that they liked, and integrated them into their own writing. Some participants clicked the ``refresh" key more than times -- that is more than 60 different writing suggestions, ``\textit{just to get a vibe}.''

\subsubsection{Even Machine-Generated Sentences Have Ramifications Beyond Themselves}

\begin{sidebar}
% \textit{%Some of the things, the inspirational stuff (quote), 
% I need to know that (machine-generated) content better as my name is attached to this document, and I need to refer to it and talk to it. %[...] I feel like I need to write it and I don't know if I trust the machine to keep my same tone and make it look like it's all me that wrote it. One thing I would love to intelligent help is [...] I'd love to see different options what it would look like.
% }~(P9)

``\textit{I need to scout out the (machine-generated) sentence. That kind of sentences has been written a million times. It is not really the point that they are trying to get. That's just a way of seeding the context. }" (P5)
\end{sidebar}

Almost all participants firmly believed that the machine-generated sentence suggestions have larger contexts and ramifications. The true intent of sentences, the philosophical stance of the author, reside in these larger contexts, ``\textit{at least two paragraphs later}". Participants shared that they needed to ``\textit{make sure this article (source article of the suggested sentence) is going where I thought was going}'', in order to assess whether the suggestion aligns with their writing. 
However, our prototype, as well as almost all sentence-level generative algorithms, does not produce such contexts, therefore simply is unable to respond to such user requests.
% When was this authored? I need some other details.

% I want to see more context of that article, or at least like... This talks about XXX, if his following argument is true, then that's probably a good thing to mention, but at the same time, the following sentence is not relevant or might not be true. 

%======== DISCUSSION ========%

\section{Prompts for Workshop Discussion}

Through the above two case studies, I draw attention to the many previously unknown complexities and nuances of ``\textit{putting ML at the service of people}" in two very different contexts, among two very different groups of users.
The intriguing responses from the clinicians and authors can serve as a point of reference for many other researchers and practitioners to interpret and discuss. The following are some possible starters of this discussion:

\begin{itemize}
    \item \textit{What issues are encompassed by the notion of ``trust"?} The cardiologists did not fully trust the DST due to at least three concerns: 1) ethical concerns about making individual-patient decisions based on population statistics, 2) the subjectivity that ML carries, and 3) the impractical notion of ``now" in ML predictions. The writers were hesitant to trust the ML suggestions for different yet related ethical and ramification concerns. How can we, the HCI research community, better scaffold and account for this nuanced and highly contextualized issue of trust?
    
    \item \textit{In designing and evaluating HCML systems, what are the trade-offs between building fully-functioning ML systems and building simulation-based experience prototypes?} Both of my studies used the latter approach, which allowed us to rapidly and iteratively experiment many designs and HCI issues without spending months or even years building the systems. 
    
    \item \textit{What are the trade-offs between lab and field evaluation?}
\end{itemize}

% I look forward to the workshop to discuss explicit considerations of HCML systems, and hope to contribute to the joint publication that workshop organizers intend to produce. 

\section{Acknowledgement}
The contents of this paper were developed under grants from the National Institute on Disability, Independent Living, and Rehabilitation Research (NIDILRR grant numbers 90RE5011 and 90REGE0007). NIDILRR is a Center within the Administration for Community Living (ACL), Department of Health and Human Services (HHS). The second case study was a collaboration with researchers at Microsoft Research during the author's internship there.

% Letting labels go, to really find what users want

% Not just asking what more people can do for algorithms, but ask what algorithms cannot do for human.

%The Importance of Fake ML Systems in Experience Prototyping

% The next two lines define the bibliography style to be used, and the bibliography file.
\bibliographystyle{ACM-Reference-Format}
\bibliography{writing,UXML,VAD}

\end{document}